# Observing with a space-borne gamma-ray telescope: selected results from INTEGRAL

Stéphane Schanne[1]

CEA Saclay, DSM/DAPNIA/Service d'Astrophysique, 91191 Gif sur Yvette, France

schanne @ hep.saclay.cea.fr

**Abstract**. The *Inte*rnational *G*amma-*R*ay *A*strophysics *L*aboratory, i.e. the INTEGRAL satellite of ESA, in orbit since about 3 years, performs gamma-ray observations of the sky in the 15 keV to 8 MeV energy range. Thanks to its imager IBIS, and in particular the ISGRI detection plane based on 16384 CdTe pixels, it achieves an excellent angular resolution (12 arcmin) for point source studies with good continuum spectrum sensitivity. Thanks to its spectrometer SPI, based on 19 germanium detectors maintained at 85 K by a cryogenic system, located inside an active BGO veto shield, it achieves excellent spectral resolution of about 2 keV for 1 MeV photons, which permits astrophysical gamma-ray line studies with good narrow-line sensitivity. In this paper we review some goals of gamma-ray astronomy from space and present the INTEGRAL satellite, in particular its instruments ISGRI and SPI. Ground and in-flight calibration results from SPI are presented, before presenting some selected astrophysical results from INTEGRAL. In particular results on point source searches are presented, followed by results on nuclear astrophysics, exemplified by the study of the 1809 keV gamma-ray line from radioactive $^{26}$Al nuclei produced by the ongoing stellar nucleosynthesis in the Galaxy. Finally a review on the study of the positron-electron annihilation in the Galactic center region, producing 511 keV gamma-rays, is presented.

## 1. Introduction to gamma-ray astronomy with space-borne observatories
The goal of gamma-ray astronomy is the study of the most extreme objects in the Universe, which include among others supernovae (e.g. explosions of stars after their nuclear fuel is exhausted), gamma-ray bursts (GRB, signed by the appearance of a new gamma-ray source in the sky, lasting most often for less than a minute, often related to the explosion of a very massive star), active galactic nuclei (accreting matter in a disk surrounding a central super-massive black-hole and ejecting matter at very high energies in jets orthogonal to the disk), or microquasars (which are the stellar-mass black-hole equivalent of active galactic nuclei). Those are sites with extreme physical conditions, involving intense gravitational fields, extreme electric and magnetic fields and very high temperatures. Those systems are often confined by gravity, which sometimes is fatal to them (when e.g. a force counterbalancing gravity stops acting, as e.g. in the case of a star whose nuclear fuel in its core is exhausted), in which case relativistic outflows of matter and the synthesis of new radioactive elements can take place and can be observed in gamma rays.

---

[1] on behalf of the INTEGRAL/SPI collaboration (CEA Saclay, CESR Toulouse, CNES Toulouse, GSFC Greenbelt, IFCTR Torino, IUMP Valencia, MPE München, UCB Berkeley, UCSD San-Diego).

For a better understanding of our Universe, multi-wavelength observations over the whole electromagnetic spectrum are favorable. Due to the presence of the Earth atmosphere (which absorbs most part of the electromagnetic spectrum, except in the visible and radio bands, see *Fig. 1*), the usage of balloon-borne or space-borne observatories is mandatory for astronomy in the X-ray and gamma-ray energy bands (from keV to hundreds of GeV).

The space-borne observatory INTEGRAL (for *Inte*rnational *G*amma-*R*ay *A*strophysics *L*aboratory) of the European Space Agency (ESA) covers the 15 keV to 8 MeV energy domain, and can be considered as the successor of both the Russian-French GRANAT mission (in orbit between 1989 and 1998), and the NASA Compton Gamma-ray Observatory CGRO (in orbit from 1991 to 2000), with its instruments BATSE, COMPTEL, EGRET and OSSE.

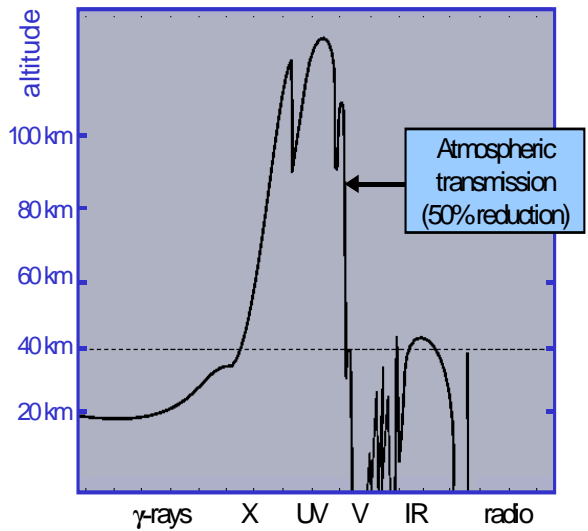 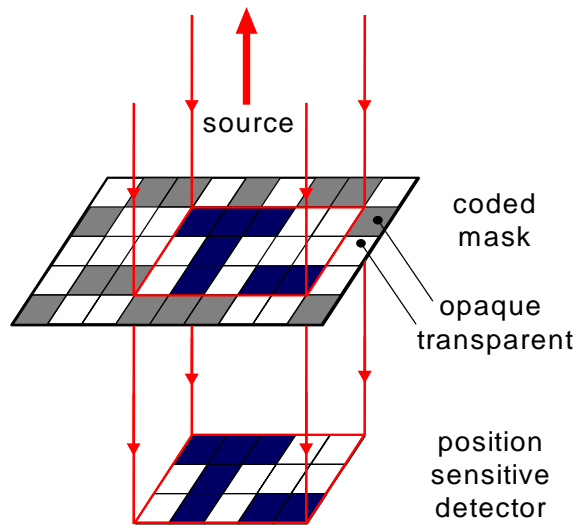

Fig. 1: Altitude at which the Earth atmosphere absorbs half of the impinging flux of electromagnetic radiation.

Fig. 2: Principle of a coded mask aperture telescope. The direction of a source is determined by the portion of the mask projected onto the detector.

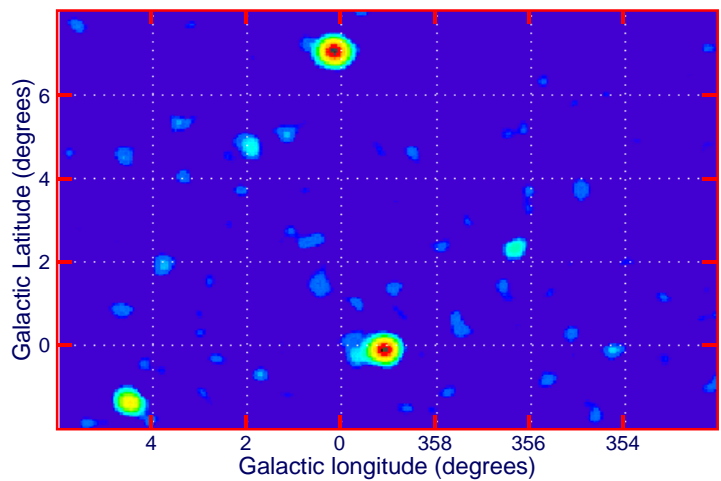

Fig. 3: The Galactic center region, as seen by SIGMA aboard GRANAT. The microquasar 1E1740-2942 is the source located near Galactic longitude=0 and lattitude=0.

Since gamma-rays can not (yet) be focussed onto a detection plane, in order to achieve a good source imaging capability, INTEGRAL uses the coded aperture mask technique (see *Fig. 2*), which is an extension of the pinhole-camera principle: a coded mask is positioned above a position sensitive detector. The mask is made of a collection of mask elements among which a part is opaque and the other part is transparent to gamma-rays. When a gamma-ray source illuminates the mask, a portion of the mask is projected onto the detection plane. The pattern of mask elements is chosen such that for any given source position in the sky, the portion of the mask pattern projected onto the detector is unique, hence determining the direction of the source. In the case of multiple sources present simultaneously in the field of view, a de-convolution technique permits the reconstruction of the positions of each of those sources from the detector image.

The previous space-borne application of the coded aperture imaging technique to gamma-ray astronomy was the French mission SIGMA ("système d'imagerie à masque aléatoire") aboard the GRANAT satellite, built by CESR Toulouse and CEA Saclay, and operating in imaging mode from 35 keV to 1.5 MeV. SIGMA permitted the discovery of the first microquasar (1E 1740.7-2942) in our galaxy [1] (see *Fig. 3*), which is a system composed of a stellar-mass black-hole accreting matter from a companion star in an accretion disk, and ejecting matter through relativistic jets which can also be observed in the radio-band trough their synchrotron radio-lobes.

## 2. Description of the INTEGRAL satellite and its scientific instruments

INTEGRAL was proposed to ESA in 1989 and a Phase-A study (proof of principle) was engaged in 1991. INTEGRAL was selected by ESA in 1993 as the next gamma-ray astronomy mission in the 15 keV to 8 MeV energy domain. In 1994 sub-contractors were invited and in 1995 the CEA was selected as project leader of the IBIS/ISGRI gamma-ray imager, while CNES was selected as project leader of the spectrometer SPI. INTEGRAL was assembled and intensely tested in ground calibrations which started in 2001. The satellite was launched in October 2002 from Baïkonur (Kazakhstan) with a Proton launcher of the Russian Space Agency. The satellite is currently performing very well and operations are approved beyond 2008.

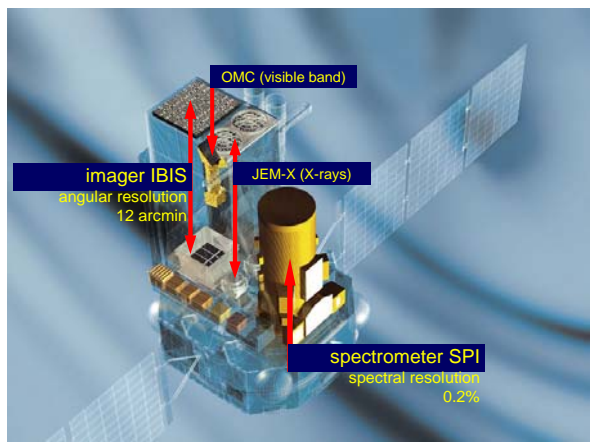
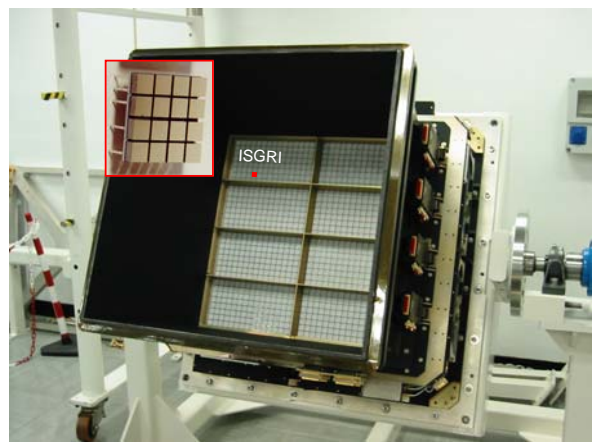

Fig. 4: Scientific instruments onboard the INTEGRAL satellite. The imager IBIS provides a very good angular resolution, while the spectrometer SPI provides excellent energy resolution. The X-ray monitor JEM-X extends the energy domain down to X-rays and the OMC permits source observations in the visible band.

Fig. 5: The ISGRI detection plane of the IBIS imager onboard INTEGRAL. The modular structure (8 modules) made of "polycells" (16 CdTe pixels) can be seen.

INTEGRAL's scientific objectives cover a wide range of astrophysics topics, like the study of :
- Nearby supernovae, if such star explosions should occur during the mission's lifetime.
- Stellar nucleosynthesis, with detection and cartography of short-, medium- and long-lived key nuclides like $^{56}$Co, $^{44}$Ti, $^{26}$Al and $^{60}$Fe and other gamma-ray line studies (like the $e^+e^-$ annihilation line).

- Structure of our galaxy, including a regular galactic plane scan for the detection of transient sources and for detailed studies of the galactic center.
- Compact Objects, identification of high energy sources, particle acceleration processes and extra-galactic gamma-ray astronomy.

The key parameters of INTEGRAL in order to reach these goals are: excellent energy resolution (for gamma-ray line studies) and excellent imaging capabilities (for the discovery of new sources). Since one single instrument could not be optimized in both parameters, two instruments (see *Fig. 4*) were built in order to separately optimize them :
- Energy resolution of about 0.2% and good narrow-line sensitivity achieved with the Spectrometer aboard INTEGRAL (SPI).
- Angular resolution better than 12 arcmin, achieved by the Imager aboard INTEGRAL (IBIS).
- The scientific performance was furthermore improved by adding an X-ray monitoring instrument (JEM-X) in order to extend the energy domain down to X-rays, and an optical telescope (OMC) in order to perform simultaneous visible band observations.

2.1. The IBIS imager and its detection plane ISGRI

The IBIS imager [2] of INTEGRAL consists of a coded aperture mask placed 3.2 m above the detection planes ISGRI and PICSIT. The ISGRI detector (see *Fig. 5*) is made of 16384 CdTe pixels, while the PICSIT detector is made of 4096 CsI pixels.

The ISRGI detector has been built under the project lead of CEA Saclay and is the first CdTe gamma-ray camera in the world (which additionally is a space borne application). The active detection area covers 2600 cm$^2$. The 16384 CdTe pixels are assembled in 8 modules, each of which is made of 128 "polycells", which are square assemblies made of 4x4 CdTe pixels. Thanks to its high Z (Cd:48, Te:52), electron/hole pairs are created with good efficiency in the inversely biased CdTe semi-conductor diode. Furthermore, thanks to its good resistivity (~$10^9$ $\Omega$ cm), the CdTe detectors work at room temperature and do not need cryogenic cooling devices.

The ISGRI detector operates in the energy band from 15 keV to ~500 keV with an energy resolution of ~8 keV (for a photon of 100 keV). The field of view is 19°x19° in the sky, and the angular resolution is better than 12 arcmin, while the point source localization accuracy is typically of 1 arcmin, and even well below 20 arcsec for bright sources. The camera is performing very well: while there were in mean ~300 dead pixels before launch, this number increased to a value of only ~400 two years after launch.

2.2. The Spectrometer SPI

The spectrometer SPI [3] is built by an international collaboration of research institutes[2], with technical leadership by the French Space Agency CNES. It is 238 cm high and has a mass of 1300 kg (see *Fig. 6*).

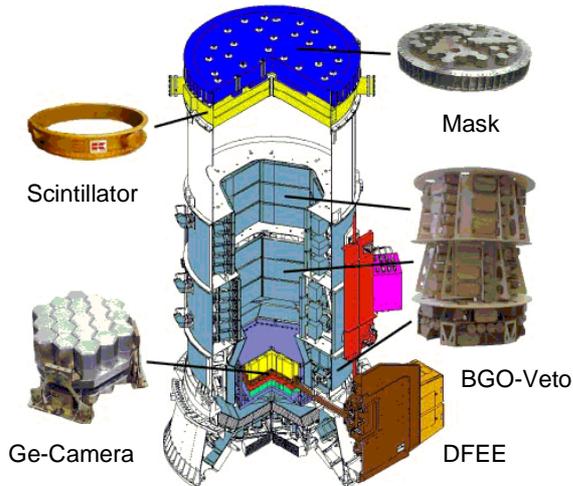
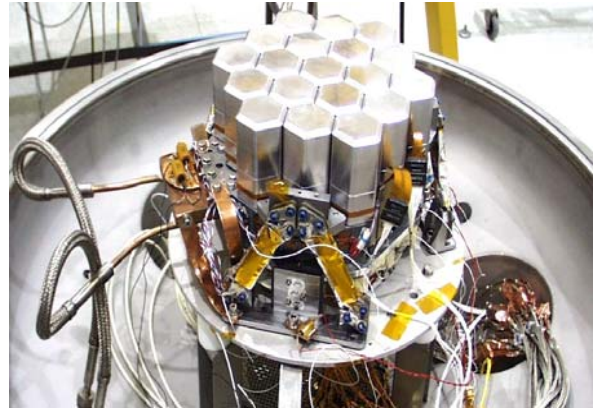

Fig. 6: The Spectrometer SPI aboard INTEGRAL, composed of the mask together with a scintillator, the anti-coincidence veto shield made of BGO crystals, the Ge detection plane and the DFEE.

Fig. 7: The Germanium camera of the Spectrometer SPI aboard INTEGRAL. 19 Ge crystals of ~1kg each are assembled together to form a detector of ~500 cm$^2$ area.

The gamma camera (*Fig. 7*) is composed of a hexagonal array of 19 Germanium detectors, covering an area of 550 cm$^2$. Each Ge detector weights about 1 kg, and is of cylindrical shape (7 cm high) with a hexagonal basis of 6 cm in size (flat-to-flat). A central bore-hole constitutes the anode of the reverse polarized diode, to which a high voltage of 4 kV is applied. Such Ge detectors provides an energy resolution of 2 keV for photons of 1 MeV. In order to reduce the background due to activation by cosmic radiation, the detectors are made of high purity Ge and the whole Ge detector array is located inside a Beryllium cryostat, in which a temperature of 85 K is maintained by an active cryogenic system, connected to a radiator, facing the anti-solar direction.

A coded mask is placed 171 cm above the SPI detection plane. The mask pattern follows a HURA structure (hexagonal uniform redundant array) and is composed of 127 hexagonal mask elements of the same size as the Ge detectors. Among those mask elements 64 are transparent elements, and 63 are opaque. Good opacity is achieved using hexagonal Tungsten blocks (W) of 3 cm thickness (99.5% opaque at 500 keV and 94.2% at 1.3 MeV). With this configuration the fully coded field-of-view of SPI covers a portion of 16° in the sky, and up to 34° of the sky are observable (with reduced sensitivity) in the partially coded field-of-view. With a mask element size of 6 cm (flat-to-flat), the angular resolution (the width of the point spread function) achieved by the SPI telescope is 2.6°.

With only 19 detectors and 127 sky elements, good imaging performance is achieved by the following trick: the same portion of the sky is subsequently observed under a slightly different telescope pointing angle, which virtually increases the number of observing detectors. To achieve this, the whole spacecraft is reoriented (a procedure called "dithering") every ~30 min using inertial wheels, according to a well defined "dithering" pattern.

For efficient background rejection the SPI Ge camera is surrounded by a massive anti-coincidence system (ACS) made of 91 BGO scintillator crystals, viewed by photo-multipliers, and a 5 mm thick plastic scintillator (PSAC) placed below the mask.

The Digital Front End Electronics (DFEE) [4], provides event timing with 100 µs accuracy, event building and classification, event rejection using the anti-coincidence veto signal, and event counting and dead-time monitoring in order to determine absolute incident photon intensities. In particular, events are classified by the DFEE (see *Fig. 8*) into Single Ge Events (SE) and Multiple Ge Events (ME) according to whether, during a 350 ns time window, a single Ge detector was hit (SE) or more than one Ge were hit (ME), which could result from Compton scattering of a photon between Ge detectors, in which case the impinging photon energy can be reconstructed by adding-up the energies deposited in the active Ge detectors. The DFEE of SPI was designed at CEA Saclay. Its core is based on a digital ASIC of 200000 logic cells built in a CMOS 0.6 μm single-event upset free and radiation tolerant technology.

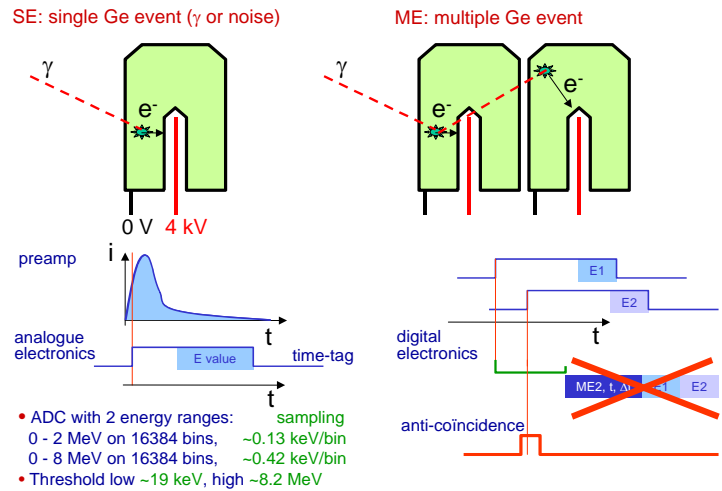

Fig. 8: Single Ge detector event (SE) and Multiple Ge detector event (ME) classified by the DFEE.

2.3. SPI ground and in-flight performance
Before launch, INTEGRAL, and in particular its Spectrometer SPI, underwent multiple dedicated ground calibrations, among which only one example will be cited here.

2.4. The SPI ground calibration
In April 2001, day and night during one month, SPI was calibrated at CEA in Bruyères le Châtel [5] (see *Fig. 9*). Detection plane efficiency and homogeneity measurements where performed over the whole energy domain with the mask un-mounted, using radioactive sources at energies below 2 MeV, and, above, a Van de Graaf high-intensity proton accelerator. Protons at 550 keV and 1720 keV were used in order to induce a nuclear resonance in a water cooled $^{13}$C target, producing excited states of $^{14}$N, which decayed producing a set of gamma-ray lines at various energies, up to 9.2 MeV. The measured detection plane efficiency [6] (see *Fig. 10*) permits an absolute spectral determination of astrophysical sources (as it has been applied to the case of the Crab).

The SPI imaging capabilities (with its mask mounted) where determined at Bruyères le Châtel at 6 different energies (from 60 keV to 2.7 MeV) using high intensity radioactive sources (up to 3 Ci) placed outside the clean room at 125 m from the spectrometer, at a distance big enough to simulate a point-like gamma-ray source. The highest energy (2.7 MeV) was obtained with a $^{24}$Na source produced on purpose in a nuclear reactor at CEA Saclay because of its short half-life of 15 h. An angular resolution of 2.55° and a point-source localization accuracy of better than 2.5 arcmin (in case of no background) was determined during those measurements.

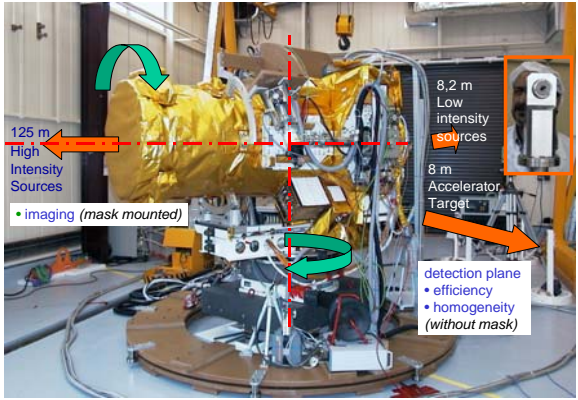

Fig. 9: The Spectrometer SPI mounted on its ground support equipment in the clean room at CEA Bruyères le Châtel (April 2001).

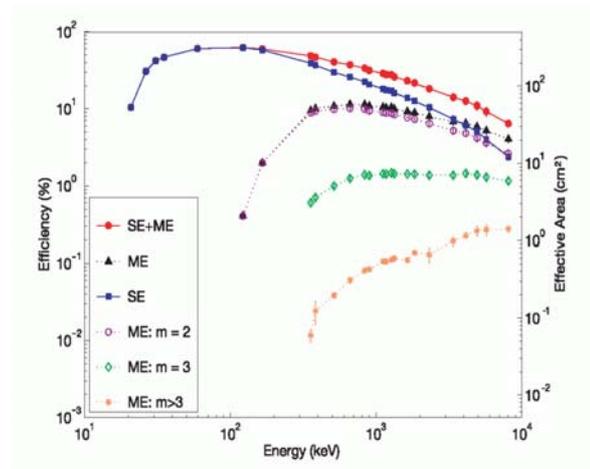

Fig. 10: Photo-peak efficiency of the SPI Ge detectors as a function of the energy of radioactive sources or nuclear gamma-ray lines used during ground calibration at CEA Bruyères le Châtel [6].

2.5. SPI in flight background

After launch during in-flight calibration [7], the first spectra acquired provide the detailed structure of the in-flight background [8], previously unknown and difficult to simulate. *Fig. 11* shows a typical background spectrum acquired by SPI in orbit. Many lines appear above a continuum spectrum. Those are due to cosmic rays interaction with nuclei of the instrument, most of which Ge activation lines and activation of materials of the massive BGO shield. Primary protons produce secondary protons and neutrons, followed by prompt or delayed gamma-ray lines. The continuum is composed of the gamma-ray Compton continuum, beta decays of nuclei and high energy particle cascades (pions). Some lines increase with time due to build-up of radioactive isotopes (as the 811 keV line of $^{58}$Co and the 1125 keV of $^{65}$Zn), while others like the 511 keV line rather follow general activity tracers like the rate of events saturating the Ge detectors (>8 MeV).

The overall count rate of a Ge detector is ~1000 counts/s (< 8 MeV) among which only ~50 counts/s are not rejected by the veto of the ACS. About ~200 counts/s are Ge saturating events (> 8 MeV). The overall dead-time is ~12%. The ACS rejection is very efficient (a factor of 25 in mean, and 130 at 511 keV). It well rejects prompt radioactive decays and cascades. It is however less efficient rejecting delayed decays, and is a source of secondary neutrons produced in the BGO, resulting in the presence of spallation lines.

The SPI background counting rates are detector dependant and variable with time. In particular during Earth radiation belt passages and during solar flares, the activity is highly variable and a good background model is important in order to search for astrophysical source signals which amount to only a few percent of the detected photons.

The SPI in flight sensitivity can be deduced from the background count-rate and the efficiency measured on ground. For narrow gamma-ray lines, SPI detects with 3σ in 1 Ms a source whose flux is 5 10$^{-5}$ ph/cm$^2$/s at 511 keV and 2.5 10$^{-5}$ ph/cm$^2$/s at 1.8 MeV. Similarly SPI can detect (at 3σ during 1 Ms) a source emitting a continuum spectrum at a level of 10 mCrab up to 200 keV and at a 1 Crab level up to 4 MeV. Due to its bigger geometrical area, ISGRI has a continuum sensitivity ~3 times better than SPI at low energy (below ~200 keV) [9].

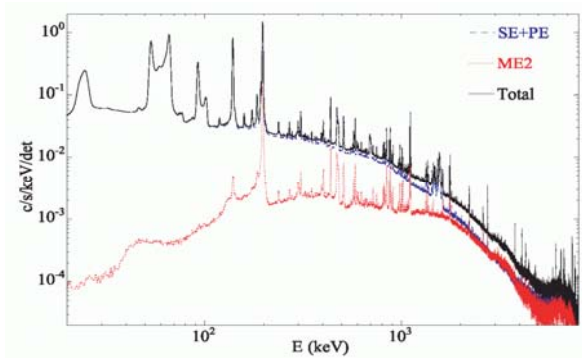

Fig. 11: SPI in flight background spectrum (with ACS active). Above 2 MeV, the ME and SE count rates are balanced and the sensitivity is greatly increase by using ME as well.

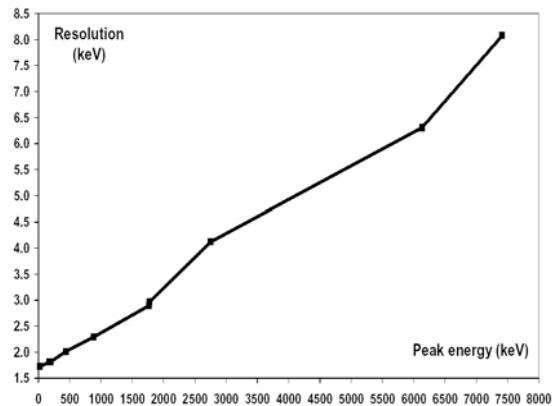

Fig. 12: SPI in flight energy resolution.

2.6. SPI in flight energy resolution and Ge crystal defect annealing

The SPI energy resolution has been determined in-flight [7] by measuring the width (FWHM) of well identified narrow background lines (see *Fig. 12*) and amounts e.g. to 0.22% at 1.117 MeV, as expected. The energy resolution is optimized by increasing the charge carrier velocities, which can be achieved by increasing the bias high-voltage (up to 4 kV) and decreasing the Ge detector temperature (down to 85 K).

With time the energy resolution degrades, due to cosmic ray interactions which displace atoms in the Ge crystal lattice and create traps for the charge carriers. These defects can be corrected by a procedure known as "annealing", which consists of heating-up the crystal: the thermal excitation permits the displaced atoms to reach again their lattice position. Annealing being a stochastic process, the longer the crystal is heated-up, the better the defects are corrected and the energy resolution recovered. From initially 36 h, the SPI in-flight annealing procedure covers now 126 h at the high temperature of 105°C. This permits less frequent annealing periods and spares observation time lost during heat-up and cool-down of the cryostat (~ 2 weeks per annealing). Annealing periods are currently carried out twice a year, fully restoring to below 3.0 keV (FWHM, at 1.78 MeV) an energy resolution which had degraded to 3.5 keV [7].

Two Ge detectors have failed since launch. Detector 2 after 14 months in flight and detector 17 after 21 months. In both cases the count-rate dropped to zero and the preamplifier input seams to have been destroyed. The exact reason is not known, and it is not so simple to reproduce the effect on ground. However both failures happened within a few weeks after an annealing took place. The annealing procedure was changed in order not to heat-up the preamplifiers anymore, but only the Ge crystals. Since then two more annealing periods took place and no other detector failure occurred. Those detector failures resulted in a loss of up to 10% in continuum- and up to 20% in line-sensitivity. The modified camera response had to be re-simulated (events which were previously classified as ME and which involve a deficient detector are now SE).

## 3. Selected INTEGRAL results

After 3 years in flight, INTEGRAL has provided numerous scientific results, among which only a very small selection will be presented here.

3.1. Operating INTEGRAL as an observatory

The orbit of INTEGRAL is highly eccentric, with an apogee of 153000 km and a perigee of 9000 km, an inclination of 51.6° at the beginning of the mission, and a revolution period of about 72 h. With this orbit provided by the Proton launcher, more than 90% of the time is spend above 40000 km (outside the Earth radiation belts, where the detectors are able to work properly) and the observing time is maximized.

Ground stations in Redu, Belgium, and Goldstone, USA, provide a permanent satellite communications link. The satellite control and health-check is provided by ESOC in Darmstadt, Germany. The scientific data are collected, preprocessed and distributed to the observing community by ISDC in Versoix, Switzerland. The actual observations are programmed by ISOC at ESTEC, Noordwijk, The Netherlands. The ability to acquire real-time telemetry from the satellite permits the ISDC to run the IBAS (INTEGRAL gamma-ray Burst Alert system), providing real-time GRB alerts.

INTEGRAL is an observatory, which means that about 75% of the observing time is offered to the world-wide community of observers, which apply for INTEGRAL observing time following an "Announcement of Opportunity" issued by ESA and evaluated by the "Time Allocation Committee". AO-4 is expected to start beginning of 2006 [10]. About 25% of the time is reserved for the teams which built the instrument and the launcher. The data are public after 1 year and distributed by ISDC together with the analysis software developed by the instrument teams.

3.2. Sky exposure

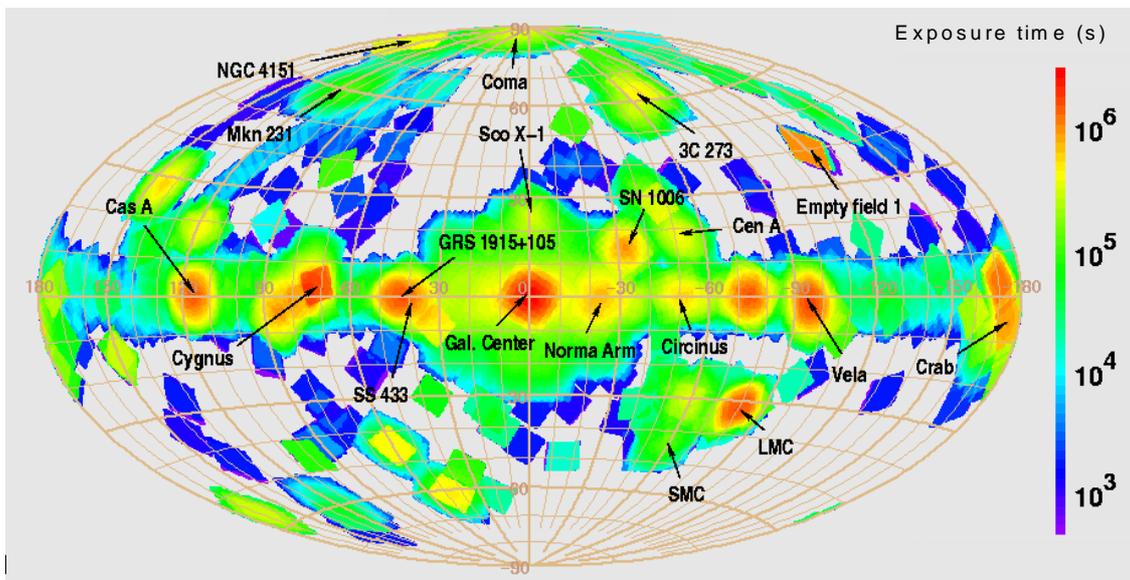

Fig. 13: Sky exposure by INTEGRAL for revolutions 1 to 200 (in Galactic co-ordinates).

The INTEGRAL observations are planned about one month in advance by the ISOC (except for targets of opportunity, where faster reaction times are required). *Fig. 13* shows the sky exposure of INTEGRAL after 200 revolutions [10]. During the first years a regular galactic plane scan was performed in order to search for new or variable sources and build diffuse emission maps. Some deep exposures covering the Galactic center region, the Cygnus or the Vela region and Galactic arms were performed, as well as exposures on selected point sources.

3.3. Example of Cygnus X-1
The archetype of a stellar mass black-hole accreting matter from a companion star via an accretion disk, Cygnus X-1, was observed very early in the mission by all 4 INTEGRAL instruments. While the OMC observes the companion star, and JEM-X the thermal emission of the accretion disk, SPI and IBIS detect non thermal emission from the inner edge of the accretion disk, resulting most likely from UV photons from the disk, which are Compton scattered on electrons falling into the black-hole, thus reaching gamma-ray energies [11]. *Fig. 14* shows the determined spectrum of Cygnus X-1 [12].

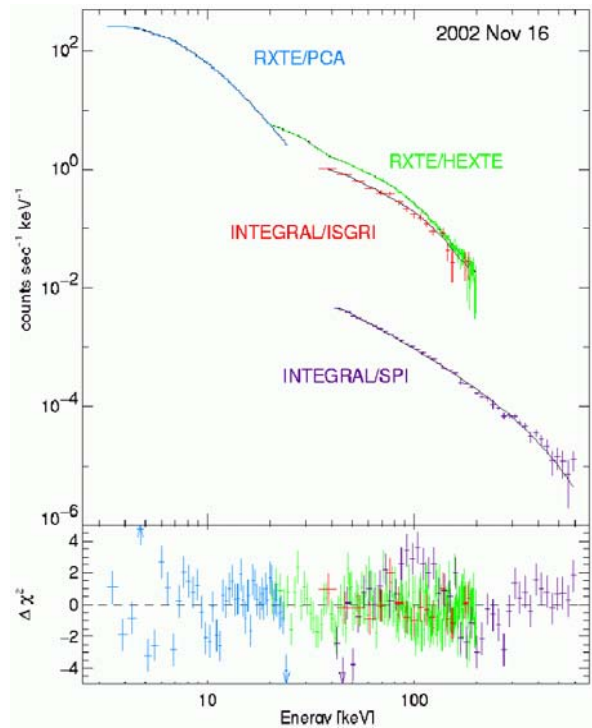

Fig. 14: Spectrum of Cyg X-1 (in counts/s) observed by ISGRI and SPI, as well as RXTE.

3.4. Example of the Crab

The Crab, the archetype of a rotating neutron star (surrounded by a strong magnetic field in which electrons undergo synchrotron acceleration producing gamma-rays), has been used for INTEGRAL as a calibration source. Using the Bruyère le Châtel calibration data, the spectrum of the Crab nebula has been determined with SPI [13] to follow a power-low of index 2.09±0.01(stat) in the 40 keV to 8 MeV band (the flux being 0.0793 ph cm$^{-2}$ s$^{-1}$ in the 50 to 100 keV band), compatible with previous determinations. The pulsar spin-period (33 ms) can be detected above 300 keV [14].

3.5. Point source observations

With 1 year of observations, mostly the Galactic plane scan data, ISGRI has found 91 point sources [15]. With 2 year of data, 158 sources have been found, as shown in *Fig. 15* in the 20-60 keV band [16]. Among these, 66 low-mass X-ray binaries, 23 high-mass X-ray binaries, 15 active galactic nuclei, 8 cataclysmic variables, 3 supernova remnants, 1 pulsar, 1 pulsar wind nebula and 41 sources of unknown type have been identified. Many of these sources have also been studies by SPI, and spectra have been determined with both ISGRI and SPI.

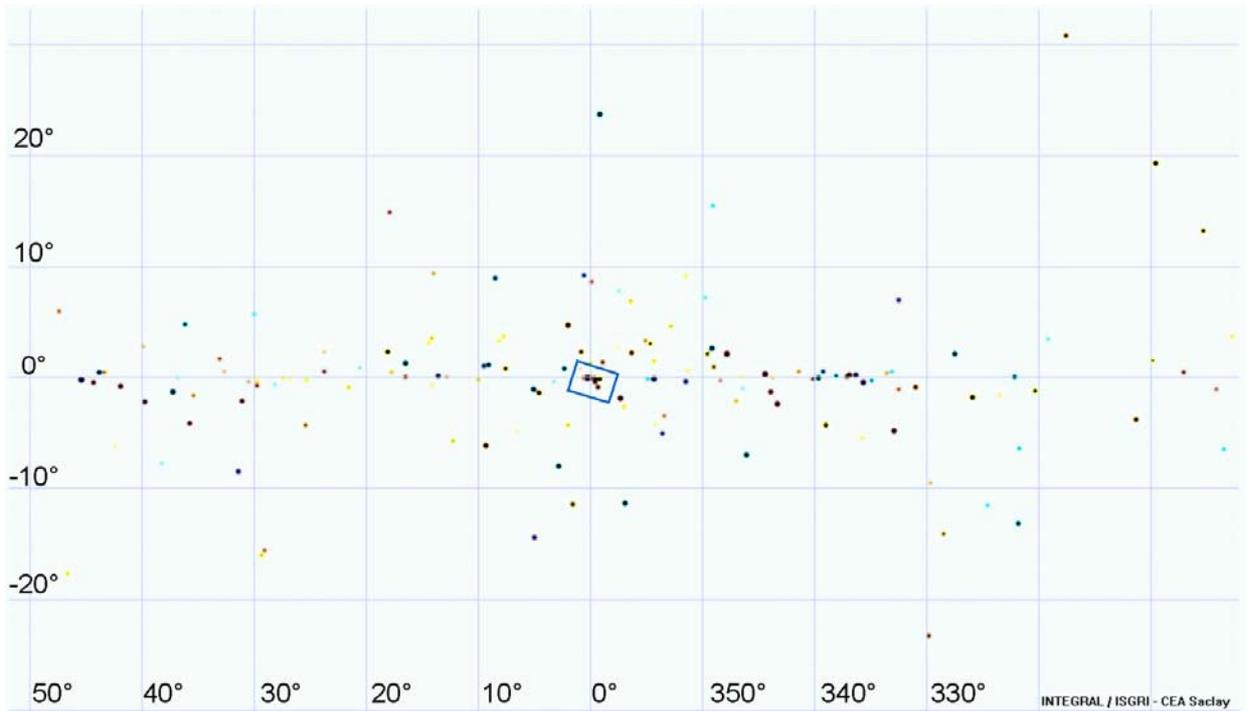

Fig. 15: Sources detected by ISGRI after 2 year of data taking (20-60 keV band), in Galactic coordinates. Color code: yellow sources have a harder spectrum, blue a softer. The box in the Galactic center region shows the zone enlarged in Fig. 16.

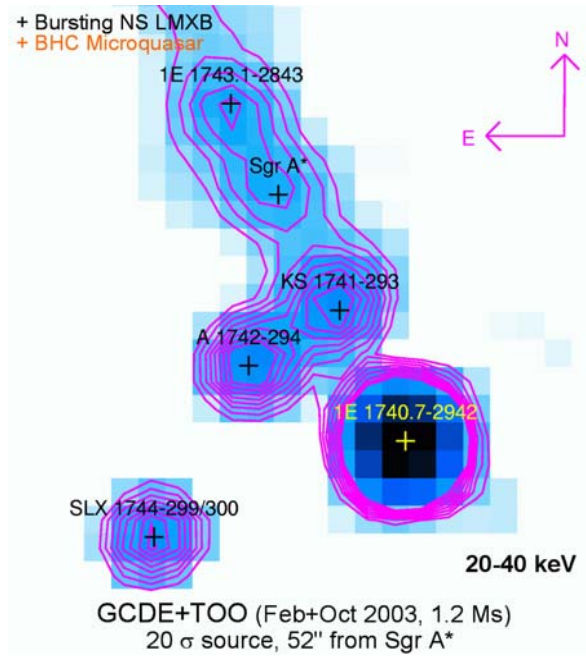

Fig. 16: Sources near the Galactic center [17], as observed by ISGRI in the 20-40 keV band, shown in right-ascension/declination coordinates (the Galactic plane goes from top/left to bottom/right). The super-massive black hole coincides with the radio source Sgr A*. ISGRI detected a new source very close to Sgr A*, maybe for the first time gamma-ray emission from Sgr A* itself.

*Fig. 16*. [17] shows a zoom on the Galactic center region, where the microquasar discovered with SIGMA in the 90's (1E1740-2942) appears as a bright source. Nearby ISGRI discovered a new source, called IGR J1745.6-2901, whose position coincides within 4 arcmin with the super-massive black-hole located at the Galactic center (Sgr A*), with a flux of 3.2 mCrab in the 20-40 keV band and 3.6 mCrab in the 40-100 keV band. The goal of further observations of this source is to see if it can be firmly

associated with Sgr A*, in particular by searching for simultaneous X-ray and gamma-ray flashes, e.g. with combined XMM and INTEGRAL observations.

### 3.6. Galactic nucleosynthesis : $^{26}$Al

Galactic nucleosynthesis, i.e. the production of new elements, still takes place today, mostly in dense media (star cores). If a synthesized radioactive nucleus is ejected from its opaque formation region into the interstellar medium before its decay, it can produce observable gamma-rays. This is the case for $^{26}$Al, believed to be produced in the core of massive stars and ejected by stellar wind (from Wolf-Rayet stars) or core collapse supernovae. With its half-life of a million year, the nucleus decays in the interstellar medium to an excited state of $^{26}$Mg, which subsequently decays and produces an observable 1809 keV photon.

In *Fig. 17* we show a preliminary detection of $^{26}$Al in the central region of the Galaxy by SPI [18] in which the counts from all detectors are added together (called "lightbucket" mode, ignoring the mask coding). The background in the 10 keV wide $^{26}$Al signal region around 1809 keV amounts to 0.35 counts/s (SE mean rate). Its time dependency has been modeled using a set of activity tracers, calibrated with high latitude data (where no $^{26}$Al signal is expected). The $^{26}$Al signal, i.e. the residual over the background model, amounts to 0.006 counts/s (mean) in the central region of the Galaxy (longitude L=0°). Since no mask information has been used here, the $^{26}$Al flux from the whole partially coded field of view is integrated at L=0°. Assuming a uniform distribution of $^{26}$Al in the central radian of the Galaxy an $^{26}$Al flux of 3.5±0.3(stat)±0.6(syst) (in units of $10^{-4}$ ph cm$^{-2}$ s$^{-1}$) can be deduced in the L=±30° region, which is compatible with the COMPTEL value of 2.8±0.2 [19]. With this flux and a Galactic $^{26}$Al distribution model, we infer [20] a total of 2.8 solar masses of $^{26}$Al in the Galaxy. The corresponding rate of core-collapse supernovae (Ib/c & II) necessary to produce this amount of $^{26}$Al is ~2 per century, in agreement with the commonly admitted rate.

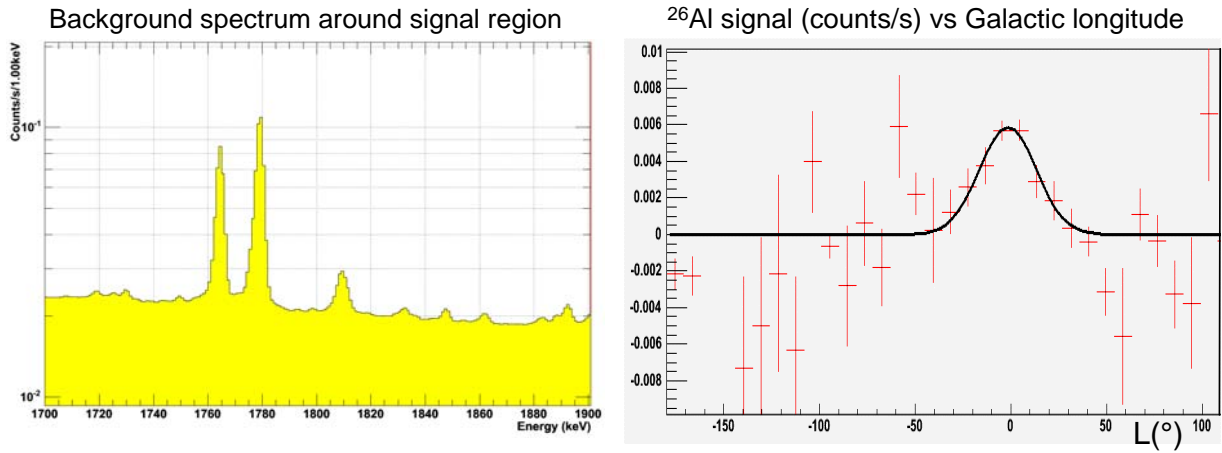

Fig 17: (Left) Background in 10 keV wide $^{26}$Al signal region. (Right) $^{26}$Al signal detected in "lightbucket" mode (SE) vs Galactic longitude (L). Error bars vary because of non uniform exposure.

*Fig. 18* shows the Galactic $^{26}$Al line detected with SPI [20] using the mask coding. Each energy bin and detector has been fit to a combination of the $^{26}$Al map determined by COMPTEL (projected through the mask) and a background model based on activity tracers. The line is intrinsically narrow [21], showing that the nuclei move with standard Galactic velocities. A line shift is detected between positive and negative Galactic longitudes, which is a hint that the $^{26}$Al nuclei follow Galactic rotation [20].

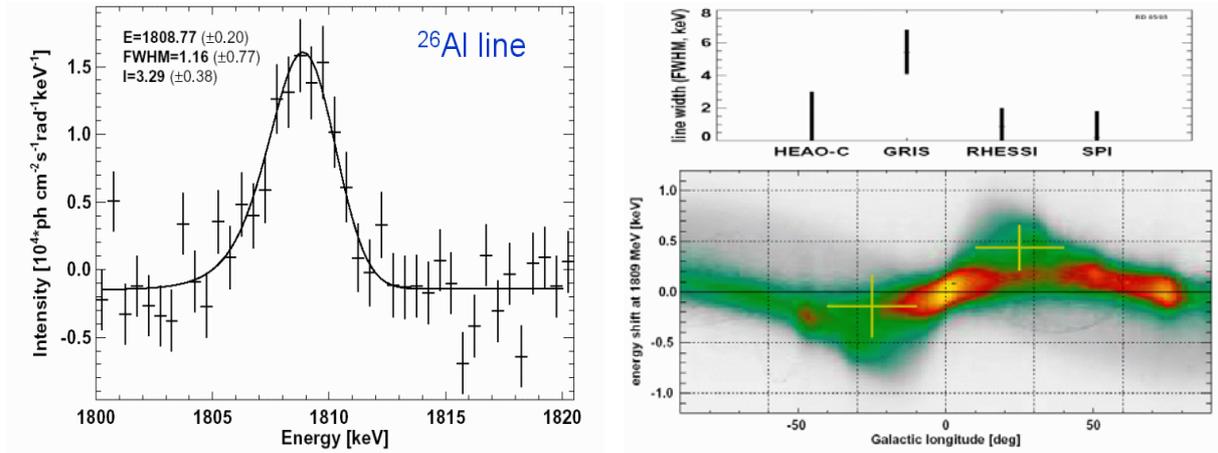

Fig. 18: (Left) $^{26}$Al line at 1809 keV, detected by SPI. (Right) The width of the line is compatible with the HEAO-C and RHESSI measurements, SPI does not confirm the GRIS result. The line is shifted between positive and negative Galactic longitude, as expected from Galactic rotation.

3.7. Galactic $e^+e^-$ annihilation radiation

One of the major results of the first two years of SPI operations is the observation of $e^+e^-$ annihilation radiation in the Galactic center region [22,23,24]. The annihilation line at 511 keV is narrow (<3 keV FWHM) and non-shifted (see *Fig. 19*).

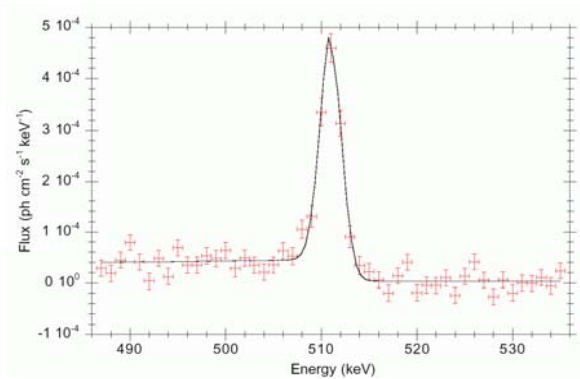

Fig. 19: 511 keV flux spectrum assuming a Gaussian distribution centered on the Galactic center.

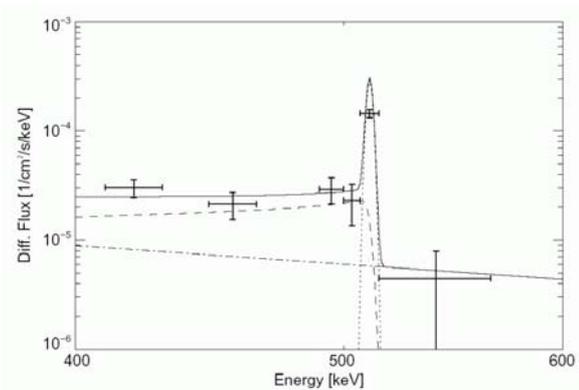

Fig. 20: Diffuse emission in the Galactic center region fitted by a 511 keV line + ortho-positronium continuum + power-law.

Positrons annihilate after thermalization, either directly or through a positronium intermediate state. As for direct annihilation, the para-positronium state produces two gamma at 511 keV, while the ortho-positronium state produces three gamma giving rise to a continuum distribution < 511 keV. A fit of these components [25] (*Fig. 20*) permits the determination of the fraction $f_{Ps}$ of $e^+e^-$ annihilation through a positronium intermediate state. The value determined from the SPI data is $f_{Ps} = (92\pm9)\%$, which confirms earlier measurements of CGRO/OSSE [26], and shows that most of the positrons annihilate after positronium formation in a warm medium.

A 511 keV emission map [27] can be determined by deconvolution of the SPI data (see *Fig. 21*). The emission is centered on the Galactic center, spherically symmetric (Gaussian of 8° FWHM). The morphologies of the 511 keV line-emission and the ortho-positronium continuum-emission coincide in shape [25], and their spatial extension is compatible in size with the Galactic bulge. This morphology is incompatible with a single point source and there is no hint that it could be produced by a combination of known point sources with a flux $> 10^{-4}$ ph/cm$^2$/s; there is also no confirmation of the positive-latitude enhancement reported by CGRO/OSSE [26]. A bulge+disk model fit to the data gives a 50σ detection of the bulge (flux: 1.05±0.06 10$^{-3}$ ph cm$^{-2}$ s$^{-1}$, e$^+$ annihilation rate: 1.5±0.1 10$^{43}$ s$^{-1}$) and a 4σ detection of the disk (flux: 0.7±0.4 10$^{-3}$ ph cm$^{-2}$ s$^{-1}$, e$^+$ annihilation rate: 0.3±0.2 10$^{43}$ s$^{-1}$). The positron annihilation rate is 3 to 9 times higher in the bulge than in the disk (while the mass ratio is only ~0.3 to 1). The e$^+$ annihilation rate in the disk is compatible with β$^+$ decays of $^{26}$Al nuclei, but the e$^+$ annihilation rate is much higher in bulge. Therefore the question about the nature of the source capable of injecting such a huge amount of e$^+$ into the Galactic bulge is raised.

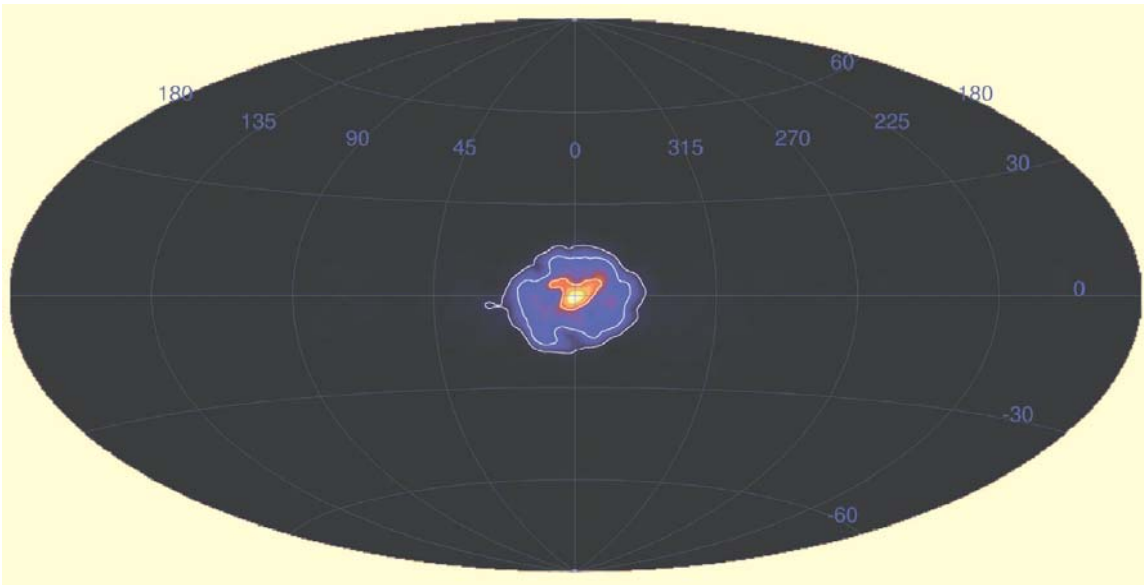

Fig. 21: Map of the 511 keV line emission obtained by .SPI (Richardson-Lucy deconvolution) [27].

Positrons can be produced by a variety of candidate sources. High energy processes like proton-proton interactions, neutralino decays or massive gravitons are excluded, since conjointly emitted high energy photons have not been observed (e.g. by CGRO/EGRET). Accreting black holes, in particular micro-quasars, could inject e$^+$, however they are not numerous enough and furthermore not permanently active. What remains are e$^+$-emitting (β$^+$) radioactive nuclei produced by stellar nucleosynthesis, among which the most prominent one is $^{56}$Co, which decays to $^{56}$Fe and emits a e$^+$ in 19% of the cases. $^{56}$Co is itself the decay product of $^{56}$Ni, synthesized massively in supernovae explosions (SN). SN of type II produce about 0.1 solar masses (Mo) of $^{56}$Ni, but all e$^+$ remain trapped inside the thick SN hydrogen envelope. SN of type Ia have a thinner envelope and produce 0.6 Mo of $^{56}$Ni, such that 3.3% of the e$^+$ escape, corresponding to 8 10$^{52}$ e$^+$ per SN Ia. An explosion rate of 0.6 SN Ia per century in the Galactic bulge would be required, if SN Ia were the dominant e$^+$ sources. However this rate has been evaluated to be 0.05 per century [28], considering the Galactic bulge as a elliptical galaxy embedded in our Galaxy and using the SN Ia-rate obtained by statistical studies of elliptical galaxies. Therefore SN Ia, considered up to now as the major e$^+$ source, are insufficient to explain the observations.

In [28,29] a new kind of e$^+$ source, namely hypernovae, have been proposed, for which the typical example is GRB 030329, one of the closest and most luminous Gamma-Ray Bursts ever observed, associated to an underlying SN of type Ic (SN 2003dh), resulting from the explosion of a rotating Wolf-Rayet star with an asymmetric ejection of matter into two opposite cones aligned with the rotation axis. Due to the ejection asymmetry the energy/mass ratio of the ejecta is much higher than in the SN Ia case and a simple model permits to compute an escape fraction of 42% of the e$^+$ produced by the 0.5 Mo of $^{56}$Ni synthesized during the explosion. In case of $^{56}$Ni mixing with the ejecta, the escape fraction could be even higher, such that a SN 2003dh-like hypernova could release up to 25 times more e$^+$ than a SN Ia. If hypernovae are the e$^+$ sources, a rate of 0.02 per century is needed. This seems in excess of very first rate estimates. However a starburst, which occurred a few million years ago in the Galactic nucleus, could have produced a sufficient number of hypernovae, and their released e$^+$ could still be annihilating today.

In the absence of a dominant astrophysical candidate, the road is open for an explanation advocating new physical processes. In the current cosmological model, the Universe is composed of 4% of baryonic matter and 23% of cold dark matter, and its geometry is rendered Euclidean by 73% of dark energy. The presence of dark matter is confirmed by studies of galactic rotation dynamics. Its nature remains a mystery for contemporary physics, and it could be made of a new kind of light particle [30,31]. Those particles could annihilate via a new intermediate boson *U* into an e$^+$e$^-$-pair and produce the observed e$^+$. Without a direct coupling to the *Z* boson, those particles could have escaped detection at LEP. Dark-matter annihilation is most likely to occur in regions of high dark-matter density, like the central region of the Galaxy [32,33]. Furthermore dark-matter rich dwarf galaxies [34] also become prominent candidate sources for 511 keV emission.

## 4. Outlook

The INTEGRAL satellite, now in orbit for about 3 years, has entered its extended mission lifetime. Despite some minor failures (2 out of 19 Ge detectors), the performance of the instrument is very satisfactory, and its operation is very smooth and reliable. The missions extension is already now approved beyond 2008. Many new and exciting results have been obtained, and many more are expected to come. IBIS has found many new gamma-ray emitting point sources, and performs spectral studies beyond 200 keV, including spectral variability studies. SPI produces broad band spectra beyond 1 MeV for many of them. Thanks to is narrow-line sensitivity, SPI performs detailed studies of nuclear astrophysics lines, such as $^{26}$Al, expected to be released by massive-star winds or explosions. SPI has also detected $^{60}$Fe [35], and with more data to come, it will be able study the morphology of its emission region, believed to be related to the one of $^{26}$Al. $^{44}$Ti is searched for in young supernova remnants, as e.g. present in the Vela region. SPI will also provide a better understanding of the morphology of the positron annihilation line in the Galactic center region, in order to better constrain the sources capable of producing the observed positrons. INTEGRAL provides also observations of transient sources, among which Gamma-ray bursts (GRB) are very exciting, as they are of cosmological origin. About one GRB is detected inside the IBIS field of view each month, and often its localization is transmitted to the observing community in very short time (10s of seconds after the event), permitting optical follow-up observations and distance/redshift determination of the host galaxy. An extraordinary "Target Of Opportunity" to study by a gamma-ray observatory like INTEGRAL would be the explosion of a nearby supernova (up to the Virgo cluster). Let's hope for such an event in the missions lifetime! With the increasing database, many more exciting results are expected. Anyone interested can apply for observing time through the ISOC "Announcements of Opportunity" [10]. Data older than 1 year are declared "public" and distributed through the ISDC [10].

**Acknowledgements**
The author would like to thank his colleagues from the SPI and IBIS collaborations for fruitful discussions, in particular B. Cordier for reviewing the manuscript. INTEGRAL is an ESA project with instruments and data center funded by ESA member states (especially the PI countries: Denmark, France, Germany, Italy, Switzerland, Spain), with the participation of Czech Republic, Poland, Russia and the USA. The SPI project has been lead by CNES, France. We are grateful to ASI, CEA, CNES, DLR, ESA, INTA, NASA and OSTC for support. 1!